\documentclass[]{aa} % for a referee version
\usepackage{graphicx}
\usepackage{txfonts}
\begin{document}
   \title{Discovery of PAHs in the Halo of NGC~5907}

%   \subtitle{I. Overviewing the $\kappa$-mechanism}

   \author{Judith A. Irwin\thanks{On leave from the
Dept. of Physics, Engineering Physics \& Astronomy,
Queen's University, Kingston, 
Canada, K7L~3N6}
          \and
          Suzanne C. Madden%\inst{1}
          }
          %\inst{1}
          %\fnmsep
   \offprints{J. Irwin}

   \institute{Service d'Astrophysique, CEA/Saclay,
              L'Orme des Merisiers, 
              91191 Gif-sur-Yvette, France\\
              \email{irwin@astro.queensu.ca}
%         \and
%             \email{smadden@cea.fr}
             }

   \date{Received  ; accepted  }

   \abstract{
We have used sensitive archival data from the Infrared Space
Observatory (ISO)
to make maps of the edge-on low SFR galaxy, NGC~5907, in 6 different MIR
bands: LW2, LW5, LW6, LW7, LW8, and LW10, covering the 
spectrum from 6.5 to 15.0 $\mu$m and including several narrow bands
that isolate the infrared aromatic spectral features commonly
referred to as PAHs.  Most of the MIR emission is dominated by
PAHs and it is likely that emission from VSGs contribute
only negligibly except in the broad IRAS-equivalent band.
The flux ratios
are typical of galaxies with low SFRs 
or quiesent regions within galaxies (e.g M~83)
and a very high PAH/continuum ratio is observed.
The PAH emission
follows the CO distribution and also shows some
 correlation within the
disk with the $\lambda$850 $\mu$m distribution.  However, 
the PAH emission
also reaches larger galactocentric radii than the CO and 
other correlations
suggest that the PAHs are also
more widespread.  A significant new discovery is the presence of PAHs in
the halo of the galaxy.  In the narrow bands that isolate single
PAH features, the emission shows structure similar to
high latitude features seen in other galaxies in other tracers. 
 The features extend as far as 6.5 kpc from the plane but 
scale heights of 3.5 kpc are more
typical.  The $\lambda$11.3/$\lambda$7.7 ratio also appears to increase
with distance from the major axis.
To our knowledge, this is the first time PAHs have been
 seen in the halo of an external galaxy.  
Just as significantly,
they are seen in a {\it low} SFR galaxy, suggesting that strong SNe and
winds are not necessary for these large molecules to reach high 
latitudes.

   }

   \maketitle
%
%________________________________________________________________

\section{Introduction}

  The galactic disk-halo interface is a critical region 
connecting two very different environments over a relatively
small vertical distance.  The prevalence, nature, and underlying physical
drivers of features which span this region
and extend into the halo are particularly important
since they play a role in the energy balance 
as well as the chemical evolution of a galaxy.
Moreover, knowledge of the  physical
conditions of outflowing gas in nearby galaxies can
provide important constraints on galaxy formation,
since outflows (i.e. some form of 'feedback') are crucial 
to galaxy formation scenarios
(e.g. Dekel \& Silk \cite{dekel};
Navarro \& White \cite{navarro}; Mo \& Mao \cite{mo};
Scannapieco et al. 
\cite{scannapieco}; Kay et al. 
\cite{kay}, Zhao et al. \cite{zhaoa}, \cite{zhaob};
Sommer-Larsen et al. \cite{sommer-larsen};
Monaco \cite{monaco}; Marri \& White \cite{marri}).  Nearby edge-on galaxies
therefore, provide an important local laboratory within which 
such flows can be examined in some detail.

Evidence for outflows includes optical emission line
cones from nuclear starbursts (e.g. Ohyama et al. \cite{ohyama};
Veilleux et al. \cite{veilleux} and others),
H$\alpha$ disk-halo filaments (Rossa \& Dettmar \cite{rossa}),
X-ray halos (Wang, Chaves, \& Irwin \cite{wang};
Strickland et al. \cite{stricklanda}, \cite{stricklandb}), and
supershells/bubbles
in HI (e.g. Spekkens, Irwin, \& Saikia \cite{spekkens})
and other bands (L{\'i}pari et al. \cite{lipari}).
Dust in halos has been
observed in absorption against the background stellar continuum
(Sofue \cite{sofue87}; Ichikawa et al. \cite{ichikawa}; 
Alton et al. \cite{alton99}; 
Howk \& Savage \cite{howk}; Thompson et al. \cite{thompson}),
although this only probes halo dust fairly close to the disk.
More recently, however, IR, mm, and sub-mm emission from 
dust
has also been detected in halos
(Alton et al. \cite{alton98}; Neininger \& Dumke \cite{neininger}; 
Brar et al. \cite{brar03}; Popescu et al. \cite{popescu04};
Tuffs \& Popescu \cite{tuffs}).
The physical conditions in such halos are, however, not well
understood, especially those which
allow the survival of dust in hot rarefied halo environments.
Indeed, the conditions of
dust survival in halos may be similar to those of elliptical galaxies
(Tsai \& Mathews \cite{tsai}; Xilouris et al. \cite{xilouris}).  

The mid-infrared (MIR) wavelength regime has the potential  
to characterise the dust properties in halos.  The sensitive 
ISOCAM detector (Cesarsky et al. \cite{cesarsky}) 
on board the Infrared Space 
Observatory (ISO) satellite (Kessler et al. \cite{kessler})
provides necessary spatial resolution and, with an appropriate selection
from the vast assortment of filters
(from 4 to 18 $\mu$m), it is possible to discriminate 
between contributing sources of dust. By now, almost a decade after the 
launch of ISO, we are beginning to understand the properties of the 
various components in the MIR regime and what they are tracing in 
galaxies. 
ISOCAM surveys of galaxies (Rigopoulou et al. \cite{rigopoulou};
Dale et al \cite{dale00}; Laurent 
et al. \cite{laurent}; Roussel et al. \cite{roussela}; 
Sturm et al. \cite{sturm02}) 
and detailed studies of individual 
galaxies (e.g. Galliano et al. \cite{galliano03}; 
Vogler et al. \cite{vogler}) have 
brought deep insight into the extragalactic MIR picture.  Among
the most prominent of the MIR features in galaxies are
the carriers of the unidentified infrared bands, one model for these 
being the PAHs (polycyclic aromatic hydrocarbons)\footnote{The carriers
of the MIR bands are not certain, but in this paper, we adopt the
PAH nomenclature for discussion and comparison with other authors.} 
and small hot (typically a few hundred K, see Sect.~\ref{spectrum})
grains.  

In this study, we use
the very sensitive previously unpublished observations of NGC~5907 in
the ISO database.  This has allowed us
to take advantage of an
extraordinary number of filters that
were used to observe both along the disk as well as into the 
halo region, where we make clear detections far from the plane.
Combining these data, we are 
able to delve into the MIR properties of the halo.
This paper is organized as follows.
In Sect.~\ref{ngc5907} we review
the previous observations of NGC~5907, in
Sect.~\ref{observations_reductions} we discuss the observations and data
reduction, Sect.~\ref{spectrum} reviews the various contributors to the 
MIR observing bands, Sect.~\ref{results} presents the results for
both the disk and halo regions, Sect.~\ref{discussion} provides the
discussion and the conclusions are given in Sect.~\ref{conclusions}.

\subsection{NGC~5907}
\label{ngc5907}

NGC~5907 (Table~\ref{basic_parameters}),
at a distance of 11 Mpc (1 arcsec = 53.3 pc), 
is one of the largest edge-on
systems in the sky.  Its nucleus is an HII region type 
 and no
 nuclear radio source has been reported (Ulvestad \& Ho
\cite{ulvestad}) in spite of some efforts to find one. 
Non-circular molecular gas motions near the nucleus 
have been interpreted in
terms of a bar (Garcia-Burillo et al. \cite{garcia-burillo97};
Garcia-Burillo \& Guelin \cite{garcia-burillo95}); however,
there is no evidence for a bar in the near infra-red (Jarrett et al.
\cite{jarrett}).  The CO distribution is centrally peaked 
(Dumke et al. \cite{dumke97})
with a distribution suggestive of rings or spiral arms 
(Sofue \cite{sofue94};
Dumke et al. \cite{dumke97}).

\begin{table*}%{lcc}
%\tabletypesize{\scriptsize}
%\rotate
\caption{Basic Galaxy Parameters\label{basic_parameters}}
%\tablewidth{0pt}
\centering
\begin{tabular}{lcc}
\hline
\hline
{Parameter\hfill} & {NGC~5907} & {M~83}\\
\hline
RA (J2000)$^\mathrm{a}$ (h m s)   & 15 15 53.69 &\\
DEC (J2000)$^\mathrm{a}$ 
($^\circ$ $^\prime$ $^{\prime\prime}$)   & 56 19 43.9 &\\
V$_{hel}$$^\mathrm{a}$ (km s$^{-1}$) & 667 & \\
Distance (Mpc) & 11$^\mathrm{b}$ & 4.5$^\mathrm{c}$\\
Morph. Type$^\mathrm{a}$ & SA(s)c: sp     HII: & SAB(s)c; HII \\
Incl. ($^\circ$) & 86.5$^\mathrm{d}$ & 27$^\mathrm{e}$ \\
Major $\times$ Minor Axis Diameters$^\mathrm{a}$ 
($^\prime$ $\times$ $^\prime$)
& 12.77 $\times$  1.40 &
12.9 $\times$  11.5\\
Major Axis Diameter (kpc) & 41 & 17 \\
$f_{60}/f_{100}$$^\mathrm{f}$ & 0.24 & 0.54 \\
L$_{IR}$$^\mathrm{f}$ ($L_\odot$) 
& 7.1 $\times$ 10$^9$ & 1.4 $\times$ 10$^{10}$ \\
T$_{dust}$$^\mathrm{g}$ 1-Comp ($\lambda^{-1}$) & 26 K & 36\\
T$_{dust}$$^\mathrm{g}$ 1-Comp ($\lambda^{-2}$) & 22 K & 29 \\
T$_{dust}$$^\mathrm{g}$ 2-Comp ($\lambda^{-2}$) & 13, 26 K &\\
T$_{dust}$$^\mathrm{g}$ 2-Comp ($\lambda^{-1.5}$) & 13, 28 K &\\
M$_{dust}$ (M$_{\odot}$) & 
6.0 $\times$ 10$^7$$^\mathrm{h}$, 3.2 $\times$ 10$^7$$^\mathrm{i}$  &
1.8 $\times$ 10$^6$$^\mathrm{i}$\\
$\Sigma_{dust}$$^\mathrm{j}$ (M$_\odot$ kpc$^{-2}$)  &
7.9$\,\times\,10^{5}$ & 2.6$\,\times\,10^{5}$ \\
SFR (M$_\odot$ yr$^{-1}$)  & 2.2$^\mathrm{h}$, 
1.2$^\mathrm{k}$ & 2.4$^\mathrm{j}$\\
SFR/A$^\mathrm{l}$ (M$_\odot$ yr$^{-1}$ kpc$^{-2}$) & 
1.7$\,\times\,10^{-3}$,
9.0$\,\times\,10^{-4}$ & 1.1$\,\times\,10^{-2}$ \\
\hline
\end{tabular}
\begin{list}{}{}
\item[$^\mathrm{a}$]
From the NASA Extragalactic Database (NED). This center is
the 2MASS center which agrees with the dynamical center 
(Dumke et al. \cite{dumke97}) to within
$\Delta\,$RA = $0.8^{\prime\prime}$ and 
$\Delta\,$DEC = $0.3^{\prime\prime}$.
\item[$^\mathrm{b}$] Sasaki \cite{sasaki}.
\item[$^\mathrm{c}$] Thim et al. \cite{thim}.
\item[$^\mathrm{d}$] Garcia-Burillo et al. \cite{garcia-burillo97}.
\item[$^\mathrm{e}$] From the major and minor axis diameters and a 
thin disk assumption.
\item[$^\mathrm{f}$] Flux ratio or FIR luminosity, using
the 60 and 100 $\mu$m IRAS fluxes
from Sanders et al. \cite{sanders}.  $L_{IR}$ 
allows for extrapolations below 40 and above 120 $\mu$m and adjusts
to the distance given in this table.  
\item[$^\mathrm{g}$] Dust temperature, assuming the emissivity laws shown, for
one or two component fits (Bendo et al. \cite{bendo03}). For M~83, a 2-component
model does not improve the fit. The second line for
the 2-component model gives the fit from
Alton et al. (\cite{alton04}).
\item[$^\mathrm{h}$] From the
 2-component dust model of Misiriotis et al. \cite{misiriotis}.
\item[$^\mathrm{i}$] From Bendo et al. (\cite{bendo03}) for fluxes within a 
135$^{\prime\prime}$ aperture, assuming a $\lambda^{-2}$ emissivity
and adjusting to the distances in this table.
\item[$^\mathrm{j}$] Dust surface density, using the dust masses 
in this table (from the adjusted Bendo et al. data) and their aperture sizes.
 \item[$^\mathrm{k}$] From L$_{IR}$ of this table and the calibration of
Kennicutt \cite{kennicutt}.
 \item[$^\mathrm{l}$] SFR per unit area, from values in the immediately preceding row
and the optical major axes given in this table.
\end{list}
\end{table*}

NGC~5907 has been an important target in
searches for faint stellar halos since it has no
appreciable bulge and was originally thought to
be isolated.   Sackett et al. \cite{sackett} 
first reported the
detection of an R-band halo to a distance of 6 kpc around this
galaxy,  following which 
V and I band extraplanar
light was detected by Lequeux et al. \cite{lequeux}, as well as J and K
band halos by
James \& Casali \cite{james} and Rudy et al. \cite{rudy}.

NGC~5907 is, in fact, a member of
the 396th Lyon Group of Galaxies (LGG~396)
of which there are 4 identified members:  NGC~5866, NGC~5879, NGC~5907,
and UGC~9776 (Garcia \cite{garcia}).  These members are at large
separations  (Table~\ref{group_parameters})
and are unlikely to be interacting 
with NGC~5907 now.  However, 
the more recent detection
of a nearby companion, PGC~54419, at a projected
distance of only 36.9 kpc and velocity separation of 
$\Delta\,V\,=\,45$ km s$^{-1}$ (Shang et al. \cite{shang})
indicate that NGC~5907 is not as isolated as previously thought.
Moreover, their additional
discovery of a large optical ring 
around the galaxy 
indicates that some interaction has indeed occurred.  A
pronounced HI warp 
 (Sancisi \cite{sancisi}) and some evidence for
an optical warp (Morrison et al. \cite{morrison}) are consistent with this. 
 The ring and nearby companion are faint
in comparison to NGC~5907; 
 the luminosity of
the ring is $\le\,1.2$\% of NGC~5907 and the mass of the second companion is
$0.5$\% of the mass of NGC~5907 (Shang et al. \cite{shang}).
Dynamical modelling of the ring indicates
that it was formed at least 0.8 Gyr ago via the destruction of 
a satellite of mass $2\,\times\,10^8$ M$_\odot$ 
(Johnston et al. \cite{johnston}, who adopted
a slightly larger distance to the galaxy).
Follow-up observations showed that the ring is highly asymmetric and
suggests that
NGC~5907 does not after all have a faint extended stellar 
halo as originally envisioned
(Zheng et al. \cite{zheng}).  A near-infrared 
(3.5 - 5 $\mu$m) search also failed to show
evidence for a halo associated with NGC~5907 (Yost et al. \cite{yost})
and HST observations find fewer bright giants in the halo region
than would be expected from a halo with standard dwarf-to-giant
ratios (Zepf et al. \cite{zepf}).
Recent 2MASS K$_s$ band observations of NGC~5907 
(Bizyaev \& Mitronova \cite{bizyaev})
provide a stellar scale height of 
$z_0\,=\,0.49$ kpc with an isothermal assumption [i.e.
$\rho(z)\,\propto\,sec^2(z/z_0)$].

\begin{table*}
%\tabletypesize{\scriptsize}
%\rotate
\caption{Galaxy Group Membership$^{\mathrm{a}}$\label{group_parameters}}
\centering
\begin{tabular}{lccccc}
\hline\hline
{Galaxy Name\hfill} & {Velocity} & 
{RA} & {DEC} &{Separation$^{\mathrm{b}}$}& 
{Separation$^{\mathrm{b}}$}\\
 & (km s$^{-1}$) & (h m s) & ($^\circ$ $^\prime$ $^{\prime\prime}$) &
(Arcmin) & (Radii) \\
\hline
NGC~5907 & 667 & 15 15 53.69 
& 56 19 43.9
& --- \\
NGC~5866 & 672 & 15 06 29.56   & 
55 45 47.9 & 85.8 & 13.4\\
NGC~5879 & 772 & 15 09 46.77  & 
57 00 00.7  & 64.5 & 10.1 \\
UGC~9776 & 833 & 15 13 07.44
& 56 58 07.3
& 44.7 & 7.0 \\
PGC~54419 & 712 & 15 14 48.04
& 56 27 15.4
& 11.8 & 1.8\\
\hline
\end{tabular}
\begin{list}{}{}
\item[$^{\mathrm{a}}$]
 Members of the Galaxy Group LGG~396 (Garcia \cite{garcia}) as well
as PGC companion found by Shang et al. \cite{shang}.  All values are 
taken from NED.
\item[$^{\mathrm{b}}$]
 `Separation' indicates the distance of the group member from
the center of NGC~5907 in units of arcmin and in units of the radius
of NGC~5907 (see Table~\ref{basic_parameters}).
\end{list}
\end{table*}

As for a gaseous halo, 
no high latitude ionized
gas has yet been detected
(Rand \cite{rand}) nor did Howk \& Savage \cite{howk} detect any
extraplanar dust in optical absorption.
 Dumke et al. \cite{dumke97}
 find a 
CO FWHM of 400 pc and an HI FWHM of
1.5 kpc, although the warp makes this latter value uncertain.
The most extended component
appears to be
the radio continuum (Dumke et al. \cite{dumke00}), for which
an exponential thick disk component is found with a scale height
of 1.5 kpc.  An 850 $\mu$m map has recently been published by
Alton et al. (\cite{alton04}) who find an extraplanar exponential
scale height of 0.11 kpc.

ISO 60, 100, and 180 $\mu$m fluxes as well as a 12 $\mu$m image
have been published by Bendo et al. (\cite{bendo02}), and 
Bendo et al. (\cite{bendo03})
also present temperature fits
to the data.  
More recently, Alton et al. have also fit a two-component dust
model to NGC~5907 which includes newer 850 $\mu$m data from
observations using the Submillimetre Common Use Bolometric Array (SCUBA)
on the James Clerk Maxwell Telescope.
These temperatures
are presented
in Table~\ref{basic_parameters}.  An earlier 2-component dust temperature
model of 18 K and 54 K had been reported by Dumke et al. (\cite{dumke97})
who observed the 1.2 mm dust component.  Note that two-temperature
models are simply a rough approximation to the range of temperatures
that likely exist in this galaxy.
 Other
parameters for NGC~5907 are also listed in Table~\ref{basic_parameters}
including the star formation rate (SFR) which is quite low
(cf. SFR = 3.8 $M_\odot\,yr^{-1}$ for the quiescent galaxy,
NGC~891, Popescu et al. \cite{popescu00}).  In Table~\ref{basic_parameters}
we also provide 
comparative data on the nearby galaxy, M~83, whose spectral
energy distribution (SED) has recently been studied in some detail
(Vogler et al. \cite{vogler}) and with which we compare our NGC~5907 results.

%__________________________________________________________________

\section{Observations \& Data Reduction}
\label{observations_reductions}

\subsection{Observations}
\label{observations}

Data on NGC~5907 were available in filters, denoted
LWn, where LW indicates the long wavelength array 
of the ISOCAM detector 
and n specifies the wavelength band.
Details of the data acquisition and resulting maps are
provided in Table~\ref{observing_map}.  As we prefer to designate
the bands according to their wavelength, we use the following
nomenclature:
LW5, LW6, LW7, LW8, LW2 and LW10 are referred to as
6.8N, 7.7N, 9.6N, and 11.3N, 6.7W, and 12W, respectively,
where the number represents the central wavelength of the filter
and N/W refers to a 'narrow'/'wide' band.

\begin{table*}
\caption{Observing \& Map Parameters\label{observing_map}}
\centering
\begin{tabular}{lcccccc}
\hline\hline
{Parameter\hfill} 
& {6.8N (LW5)}& {7.7N (LW6)}& 
{9.6N (LW7)}& {11.3N (LW8)} &{12W (LW10)} &
{6.7W (LW2)} \\
\hline
Central Wavelength$^\mathrm{a}$ 
($\mu$m) & 6.8 & 7.7 & 9.6 & 11.3 & 12.0 & 6.7\\
Frequency Range$^\mathrm{a}$ ($\mu$m) 
& 6.5 - 7.0 & 7.0 - 8.5 & 8.5 - 10.7 & 10.7 - 12.0 
& 8.0 - 15.0 & 5.0 - 8.5 \\
Observing Mode$^\mathrm{a\,b}$ 
& CAM01  & CAM01 & CAM01 & CAM01 & CAM01 & CAM03\\
TDT No.$^\mathrm{b}$
& 09801401 & 09801401 & 09801401 & 09801401 & 26902911 & 15601301 \\
             &        &       &       &       &       & 15601202 \\
             &        &       &       &       &       & 15601103 \\
             &        &       &       &       &       & 15601004 \\
Pixel Field of View (arcsec) & 6.0 & 6.0 & 6.0 & 6.0 & 1.0 & 6.0\\
PSF (FWHM) (arcsec) & 7.2 & 7.2 & 7.2 & 7.2 & 4.2 & 7.2 \\
Date of Observations & 23 Feb 96 &23 Feb 96 & 23 Feb 96& 23 Feb 96 
& 12 Aug 96 & 21 Apr 96\\
No. of On-Source Pointings$^\mathrm{c}$ 
& 3 & 3 & 3 & 3 & 12 & 4 \\
Mean No. Frames per Pointing & 16 & 16 & 17 & 17 & 15 & 248 \\
Integration Time per Frame (s) &  
10.0804 & 10.0804 &10.0804 &10.0804 & 1.12 & 10.0804\\
Total On-Source Time$^\mathrm{d}$ (min) 
& 8.1 & 8.1 & 8.6 & 8.6 & 3.4 & 167\\
Sky Coverage$^\mathrm{e}$ (arcmin$^2$) 
& 8.2 $\times$ 3.2 & 8.2 $\times$ 3.2 &
8.2 $\times$ 3.2 & 8.2 $\times$ 3.2 & 2.8 $\times$ 1.6  & 8.4 \\
Calibration Error$^\mathrm{a}$ (\%)
& 7.5  & 6.4 & 6.7 & 5.4 & 3.9 & 3.3 \\
Sky Level$^\mathrm{f}$ (mJy pixel$^{-1}$) 
& 1.39 & 3.28 & 7.19 & 10.27 & 0.27 
& 1.99 \\
1$\sigma$ Sky Error$^\mathrm{g}$ (mJy pixel$^{-1}$) 
& 0.24 & 0.15 & 0.28 & 0.30 & 
0.015 \\
%nb judith the above is truly a fwhm not a sigma
$\sigma_{Rms}$$^\mathrm{h}$ (mJy pixel$^{-1}$) 
& 0.44 & 0.17  & 0.15 & 0.24 & 0.062 \\
Transient Rms$^\mathrm{i}$ (mJy pixel$^{-1}$) 
& 0.35 & 0.24 & 0.16 & 0.23 & 0.044 \\
$\sigma$$^\mathrm{j}$ (mJy pixel$^{-1}$) 
& 0.61 & 0.33  & 0.36 & 0.45
& 0.077 & 0.025$^\mathrm{k}$ \\
$\sigma$$^\mathrm{l}$ (mJy arcsec$^{-2}$) 
& 0.017 & 0.0092 & 0.010 &
0.013 & 0.077 & 0.00069 \\ 
Dynamic Range$^\mathrm{m}$ & 20 & 72 & 28 & 41 & 6.9 & 702 \\
\hline
\end{tabular}
\begin{list}{}{}
\item[$^{\mathrm{a}}$]  Blommaert et al. \cite{blommaert}.  
\item[$^{\mathrm{b}}$]
CAM01 is the general observing mode for photometric imaging.
CAM03 is a beam-switching mode for photometric imaging. 
TDT no. is a unique number that identifies the observation.
\item[$^{\mathrm{c}}$]
 This is the number of different pointing positions on-source; for
some pointings, the fields overlap.
\item[$^{\mathrm{d}}$] Total integration time per on-source pointing
times the number of pointings.  Time on sky not included.
\item[$^{\mathrm{e}}$]
 Total area of sky covered after registration of the different
on-source pointings.
\item[$^{\mathrm{f}}$]
Median 
value of the histogram of all sky pixel values which were subtracted
from the image. %judith for lw6,7,and 8, i just measured from the plots
%since i hadn't recorded the values
\item[$^{\mathrm{g}}$] 1$\sigma$ width of the histogram of sky pixel values 
which were used to determine the sky background median.
 %judith for lw5,6,7,8 (ie.
%all, i just measured from the plots -- thus somewhat approximate -- sky
%sky_lw5.ps etc to see the plots
\item[$^{\mathrm{h}}$]
 Median $\sigma_{Rms}$ over the field, 
where the $\sigma_{Rms}$ is the per 
pixel error from the dispersion in the signal 
carried through by the software throughout the data reductions.
\item[$^{\mathrm{i}}$]
 Median rms over the field, where the rms is the error due to 
transients.
\item[$^{\mathrm{j}}$]
Quadratic sum of the random errors from the immediately preceding
3 rows, unless otherwise indicated.  This error represents a typical random
error for any pixel in the map, but there are positions at which the error
is larger or smaller.  Note that the calibration error is not included.
\item[$^{\mathrm{k}}$] Median rms error over the 4 fields which constitute the
mosaic, including all applicable random errors.  For this beam-switching
observation, the above errors were factored into this error in the software
and were not calculated separately.
\item[$^{\mathrm{l}}$]
 Random error from previous row per 1 arcsec square pixel.
\item[$^{\mathrm{m}}$] Maximum of map divided by the median Random error.
\end{list}
\end{table*}

The first 4 data sets were taken on 23 Feb 1996 in standard raster mode
(CAM01) in the narrow bands, 6.8N, 7.7N, 9.6N, and 11.3N.
 Each observation had 3 different on-source
pointings for a total field which provided full coverage of the 
south-east part of the galaxy as well as
sufficient coverage far from the emission for sky subtraction.
For each pointing, a series of frames were taken over a period of time
 in a 'temporal block'.  The
approximate number and duration of the 
frames are given in Table~\ref{observing_map}.

The 5th data set was taken on 12 Aug 1996 (12W). 
These data were taken in microscan raster mode (CAM01)
with 12 observing positions in a 6 $\times$ 2 pattern
(Bendo et al. \cite{bendo02}) 
with a resulting 1$^{\prime\prime}$ pixel spacing.
In this case, 12
on-source pointings were made for full coverage of the central part of
the galaxy with sufficient sky coverage for subtraction.

The final data set, 6.7W, was obtained on 21 Apr 1996 in beam switching
mode (CAM03).  There
were 4 positions on-source:  one containing the galaxy center but 
offset to the south-west, another containing the galaxy center offset
to the north-east, one at the end of the major axis to the north and one
at the end of the major axis at the south. Thus, the final combined image does
not provide full coverage of the whole galaxy but rather samples the galaxy
at four locations.
 For each of the 4 pointings, the off-source sky positions were placed
 at a variety of position angles
with respect to that pointing.

\subsection{Reductions}
\label{reductions} 

All data were reduced using the CAM Interactive Reduction Package
(CIR, Chanial \cite{chanial}).  
First, the dark current was subtracted following
the method of Biviano et al. (\cite{biviano}) which includes a dark correction,
a second-order dark correction depending on detector temperature and
time of observation, and a short-drift correction.  Next, high glitches
due to cosmic ray (CR) hits were removed automatically via a multi-resolution
filtering technique (Starck et al. \cite{starck}) 
at a 6$\sigma$ level.  The effects
of memory on a pixel (transient effects) 
were then corrected (de-glitched) using the 
Fouks-Schubert method (Coulais \& Abergel \cite{coulais}).  
 Each data set
was then examined carefully, frame by frame, and further bad pixels 
were removed 
manually as required.  Most of these occurred in regions immediately adjacent
to pixels which had been automatically de-glitched.

For the raster mode observations (CAM01:  
6.8N, 7.7N, 9.6N, 11.3N, and 12W), 
the edges were then trimmed
(blanked), the frames were corrected for jitter, and averaged. The results
 at each pointing were then
corrected for the flat field according to the on-line
library of calibration flat fields 
(see Roussel et al. \cite{roussela}).  The various
pointings on-source were then projected onto a common grid in sky coordinates
and corrected for field lens distortion.  
 At this point, the data were calibrated to
mJy/pixel units (Blommaert et al. \cite{blommaert}).  
Finally, the sky background/foreground
was subtracted by first isolating a region of sky-only in the field
and subtracting the median of that region from the field.

For the beam switching observations (CAM03: 6.7W) we carried out
an additional stage of 
manual deglitching to account for the fact that in the off-source positions,
the CCD contained a fading memory of galaxy itself from when the array pointed
at the galaxy.  Thus, the region of the array which had previously contained
the galaxy was blanked in approximately 10 sky frames immediately after 
moving from the galaxy to the sky.
This blanking covered about 20\% of the array and left approximately
20 frames where the image had faded and blanking was no longer required. 
This additional deglitching was required only for the center-most 
two fields, and
not for the two fields at the ends of the galaxy's major axis, given the
weakness of the signal there.
Beam switching allowed for sufficient sky coverage off source that the
sky positions provided both the flat field and its distortions as well as
the sky brightness values.  Thus, after the
average on-source image was created, it was corrected
using an image that was formed from an average of the
various off-source positions.
The result was then calibrated to
mJy/pixel units. This mode of operation provides
the best sky subtraction because it does this
 pixel by pixel rather than using a single value
(the median of the sky values) for all pixels, as above.
Finally the registration and
combination of the 4 pointings, as well as
an interpolation over one bad column on the array
 was carried out outside of the CIR package.  
In the region of the galaxy in which there was overlap (see
center region of Fig.~\ref{lw2}a),
the flux levels between the two overlapping fields
was $\sim$ 4.5\%.

All images were interpolated/rotated into RA/DEC coordinates with 
1 arcsec square pixels for consistency.  
The resulting maps are shown in Fig.~\ref{lw5}a through
\ref{lw2}a.  The lowest contour, in each case, is set to 2$\sigma_{map}$,
where $\sigma_{map}$ is the rms variation calculated from a blank sky
region on the emission map itself.  The field of view is accurately
shown 
by the error maps (next section).  The results are discussed in
Sect.~\ref{results}.

\begin{figure*}
%\epsscale{.80}
%\resizebox{\hsize}{!}{\includegraphics{fig1.ps}}
\caption{
Maps of NGC~5907 in the
6.8N band showing {\bf (a)} total emission 
and {\bf (b)} the Random error map (see
Sect.~\ref{observations_reductions}).
The greyscale is
in units of $\mu$Jy arcsec$^{-2}$ and ranges from
the minimum to maximum map values.
Contours are at
0.012 (2$\sigma_{map}$),
0.018, 0.03, 0.06, 0.10, 0.20, and 0.30 mJy arcsec$^{-2}$
where $\sigma_{map}$ is the rms noise level
as measured from the emission map itself.} 
\label{lw5}
\end{figure*}

\begin{figure*}
%\epsscale{.80}
%\plotone{fig2.ps}
%\resizebox{\hsize}{!}{\includegraphics{fig2.ps}}
\caption{
As in Fig.~\ref{lw5}, but for 7.7N
and with
 contours at
0.006 (2$\sigma_{map}$), 0.01, 0.018, 0.05, 0.10,
 0.30 0.50, and 0.65 mJy arcsec$^{-2}$.   The inset shows the
emission after an estimate for the stellar contribution has
been subtracted (see Sect.~\ref{stellar_contribution}).}
\label{lw6}
\end{figure*}

\begin{figure*}
%\epsscale{.80}
%\plotone{fig3.ps}
%\resizebox{\hsize}{!}{\includegraphics{fig3.ps}}
\caption{
As in Fig.~\ref{lw5}, but for 9.6N
  with
 contours at
0.015 (2$\sigma_{map}$), 0.025, 0.035, 0.05,
 0.10, 0.18, and 0.25 mJy arcsec$^{-2}$.}
\label{lw7}
\end{figure*}

\begin{figure*}
%\epsscale{.80}
%\plotone{fig4.ps}
%\resizebox{\hsize}{!}{\includegraphics{fig4.ps}}
\caption{
As in Fig.~\ref{lw6}, but for 11.3N
with
 contours at
0.016 (2$\sigma_{map}$), 0.022, 0.038, 0.055,
 0.10, 0.20, 0.30, and 0.45 mJy arcsec$^{-2}$.}
\label{lw8}
\end{figure*}

\begin{figure*}%[htb]
%\resizebox{\hsize}{!}{\includegraphics{fig5.ps}}
\caption{
As in Fig.~\ref{lw5}, but for 12W
  with
 contours at
0.028 (2$\sigma_{map}$), 0.05,
 0.10, 0.20, and 0.35 mJy arcsec$^{-2}$.}
\label{lw10}
\end{figure*}

\begin{figure}%[h]
%\resizebox{\hsize}{!}{\includegraphics{fig6.ps}}
\caption{
As in Fig.~\ref{lw6}, but for 6.7W
  with
 contours at
0.0008 (2$\sigma_{map}$), 0.0012,
 0.002, 0.005, 0.010, 0.030, 0.10, and 0.35 mJy arcsec$^{-2}$. 
The irregular region in (b) denotes the area common to all maps.}
\label{lw2}
\end{figure}

\subsection{Error Analysis}
\label{errors}

Attention has been paid to the error budget in the final images.
The resultant error is the quadratic sum of individual errors,
consisting of a) the readout and photon noises added quadratically
and calculated by the software, denoted '$\sigma_{Rms}$' in
Table~\ref{observing_map}, b) an error on the background level
which was subtracted; this is taken as the 1$\sigma$ dispersion
in the histogram of values used to determine the sky value to
be subtracted and includes
uncertainties due to residual dark current, flat field, glitches,
remnant images, and long-term drifts, denoted '1$\sigma$ Sky Error'
in  Table~\ref{observing_map}, 
and c) the effects of the non-stabilization of the signal (memory) 
as a function of time over a temporal frame-by-frame block
at any pointing; this error has been estimated from variations in
the signal that are greater than 3$\sigma$ within a block (assuming
that such larger variations are caused by only this error) and applying
a factor which accounts for the fact that the error is worse near the
beginning of a block 
(see Galliano \cite{galliano04}, 
Roussel et al. \cite{roussela}),
denoted 'Transient Rms' in Table~\ref{observing_map}.  The quadratic
sum of these errors is referred to as the random error ('$\sigma$'
in the table) and has been computed 'per pixel', where the pixel
is the originally sized pixel given in the table, and also has been
computed 'per square arcsec', after interpolating onto a 1 arcsec
pixel grid.  We formed a total Random Error 
map (showing $\sigma_i$ for
each map point, i,
 from the quadratic sum of the $\sigma_{Rms}$ map, the
Transient Rms map, plus a constant 1$\sigma$ Sky Error value.  These
maps are shown in Figs.~\ref{lw5}b through \ref{lw10}b.
However, for
 the beam-switching observations, since we also have the sky background
errors for each pixel (rather than the same 
sky value for every pixel), all 
three sources of error were calculated in one step (thus
the missing intermediate step values in Table~\ref{observing_map}) 
and the
final Random Error map is shown in Fig.~\ref{lw2}b.

The calibration error (see Table~\ref{observing_map})
has not been included in Figs.~\ref{lw5}b through 
\ref{lw10}b since this would globally raise or lower the flux level
of the map and these maps are mainly meant to show
variations
across the fields.  However, it is included in any calculations of 
flux density.  A remaining error, that of variations in the flux
density calibration according to orbit 
($\approx$ 5\%, Blommaert et al. \cite{blommaert})
has not been included. Since 6.8N, 7.7N, 9.6N, and 11.3N observations were
all taken at the same orbital point, this error does not enter into
a field by field comparison of these 4 bands.  However, it will be
a factor in a comparison between these bands and the remaining two.

We note that the true errors cannot be determined rigorously, for example, the
memory effects of the camera are not accurately known 
 (Roussel et al. \cite{roussela}), although they can be tested for, as
described below.
Thus the quoted errors, though calculated rigorously, 
should be considered indicative.  An example is
seen in Fig.~\ref{lw7}a in which there remain three residual artifacts
to the west of the galaxy which correspond to the edges
of the three fields.  We have also, therefore, shown the lowest
contour level
in Figs.~\ref{lw5}a through \ref{lw2}a as twice the rms level on the
emission map itself, the latter being an indication of the 
pixel-to-pixel variation across a field in the region of the sky.

Since we are reporting the discovery of halo PAHs in NGC~5907
in bands, 7.7N, 11.3N, and 6.7W
(see Sect.~\ref{high_latitude_emission}), we performed some
additional tests which were designed to check whether this
high latitude emission might have been produced artificially.
These tests are as follows, foccusing on the 7.7N band.

We first examined the point spread function (PSF) whose 
FWHM values are 7.2$^{\prime\prime}$ in these
bands (Table~\ref{observing_map}).  The PSFs
at different locations on the image
in these bands have been carefully examined and
characterized by fitting 2-D gaussians 
(see Galliano \cite{galliano04}) and are,
to a high accuracy, symmetric.  Thus, the wings of the PSF should
be completely negligible 7.2$^{\prime\prime}$ beyond the projection
of the disk.  To check this, we took the 7.7N image (Fig.~\ref{lw6}),
measured its semi-major axis (245$^{\prime\prime}$), and created 
an inclined
(86.5$^\circ$) thin disk model 
(see Irwin et al. \cite{irwin99} for similar
modelling).  The resulting projected half-width perpendicular to
the plane at the center of the galaxy is 15$^{\prime\prime}$ and
less elsewhere.  This model was then used as a template to blank
all emission {\it outside} of the inclined thin disk region from the 
Fig.~\ref{lw6} image.  The resulting blanked image then contained
only emission from the disk region.  The blanked image
was then convolved with a 7.2$^{\prime\prime}$ gaussian PSF
to create an image consisting of disk plus extended emission due
to the PSF only.  This resulting map was then compared with the original
image by taking the
 ratio of the two and determine the pencentage change per pixel
that results.
  We find that, outside of
the projection of the thin disk, all ratios are $<\, 1$\% and the
mean is $<\,0.05$\%, verifying that the PSF contributes negligibly
to the halo emission.

Secondly, we compared results between using the library flat field
(as described in Sect.~\ref{reductions}) 
and using a sky flat, formed from a 
portion of the image in which the sky is visible.
The mean relative error between the two results over comparable
regions is 4.5\% and
high latitude structure, most of which is in the region in which
the sky flat could be formed,
remains present.  Moreover, in both images,
artifacts, such as those at the
edge of the field in Fig.~\ref{lw7}a, cannot be seen.

Finally, we checked further into issues related to memory.
Although the resulting map already has memory corrections in
place including blanking the first frames of any scan, we 
applied a more aggressive scheme.  This involved 
substantial blanking of the first time period when a new
scan began, determining the median of the last 5 frames of the
scan, pixel by pixel,
 and then fitting a slope to every pixel
independently to align the other frames with this median.
Some differences in structure were noted at a low level, 
however the high latitude emission remained.

%______________________________________________________________

\section{The MIR Spectrum and Contributors to the Observed Bands}
\label{spectrum}

To interpret the emission in the observing bands, it is first necessary
to understand what the various contributors are to the MIR emission.
For this purpose, we provide ISOCAM CVF 
(circularly variable filter) spectra of 
regions in two other galaxies for comparison
(see Fig.~\ref{m83_spectrum}).  The main spectrum is of the nucleus
of the nearby galaxy, M~83, and the inset is of a quiescent region
away from the starburst in M~82.
Various
features are labelled as well as the bands used in our observations.

\begin{figure*}%[h]
%\epsscale{.80}
%\centerline{\includegraphics*[scale=0.8]{fig7.ps}}
%\plotone{fig7.ps}
%\resizebox{\hsize}{!}{\includegraphics{fig7.ps}}
\caption{Comparison MIR spectrum of the central
position of M~83 
from the ISOCAM CVF observations of Vogler et al. (\cite{vogler}).  
The major features are labelled and the 
bands used in the observations of NGC~5907 are indicated at the top.
{\it Inset:} MIR spectrum of a quiescent region in M~82, away
from the central starburst,
showing modeled components, from Laurent et al. (\cite{laurent}).}
\label{m83_spectrum}
\end{figure*}

There are two main contributors to the MIR spectrum:
{\it 1)}  strong spectral peaks due to 
 very small grains or large molecules (PAHs) 
which, due to their small sizes (typically $\sim$ 0.001 $\mu$m)
and 
thus, low heat capacity, are stochastically heated, undergoing wide 
temperature fluctuations covering 100s of K from absorption of a 
single photon 
(Leger \& Puget \cite{leger}, 
Sellgren et al. \cite{sellgren}, Allamandola 
et al. \cite{allamandola}). The primary emission bands,
relevant to the wavelength range observed for NGC~5907,
 are located at 6.2, 7.7, 
8.6, 11.3 and 12.7 $\mu$m and can be seen as strong emission peaks
in the spectrum.
{\it 2)} a continuum composed of very small (of order $\sim$ 0.01 $\mu$m)
grains (VSGs) as 
modeled in our galaxy by D{\'e}sert et al. (\cite{desert}).  
Depending on their size 
and the local radiation field, these grains could be stochastically 
heated or in thermal equilibrium.  A modeled continuum is shown in the
Figure inset, increasing in strength to longer wavelength.
Finally, not shown in the figure, is a possible weak stellar contribution
that, if present, 
 would contribute at the low wavelength end of the spectrum, decreasing
at higher wavelengths.  Since we are able to make numerical estimates of
this contribution from previous observations of NGC~5907 itself, 
we deal with this separately in Sect.~\ref{stellar_contribution}.

Without doubt, the most overwhelming emission in the MIR band is
due to PAH features.  These features are ubiquitous in normal
spirals and account for almost all of the MIR energy 
(e.g. Vogler \cite{vogler},
Genzel \& Cesarsky \cite{genzel} \& references therein).
It is also now established that most normal spiral galaxies show 
qualitatively
little difference in their spectra in the MIR 
and that the spectral shape is largely independent of
star formation (Lu et al \cite{lu}).
From higher resolution Galactic observations, we know that the VSGs
(producing the dust continuum) peak 
in the nebular regions, while the PAHs 
peak outside the HII regions, in 
the photodissociation regions (PDRs) around molecular cores (e.g. 
Cesarsky et al \cite{cesarsky}, 
Verstraete et al \cite{verstraete}). 
While the PAHs emit 
profusely in the disks of galaxies, excited predominantly by UV photons, 
they are also observed in the more diffuse ISM 
(e.g. Chan et al. \cite{chan}, 
Mattila et al. \cite{mattila96}) 
where optical photons are thought to be the source 
of excitation. Optical photons are also an explanation for the presence 
of PAHs in elliptical galaxies 
(Athey et al \cite{athey}, Xilouris et al \cite{xilouris}), 
where evolved stars are the primary stellar population.  Moreover,
in a study of 5 galaxies over a range of star forming activity,
Haas et al. (\cite{haas}) find a spatial correlation between PAHs and the
cold dust distribution which is more widespread than around
SFRs alone.  

We thus expect the relative importance of
the continuum to diminish in galaxies of lower star formation rate
or in quiescent regions of galaxies.
This is indeed observed in M~83 (the increasing continuum
longward of $\lambda$14 $\mu$m at the nucleus shown in
Fig.~\ref{m83_spectrum} becomes much less
prominent in the interarm regions) as well as other known
quiescent star forming regions (Fig.~\ref{m83_spectrum} inset). 
Thorough modeling of ISOCAM CVF spectra
using Lorentizian profiles further supports the dominance of
the PAH bands, greatly reducing the need for a strong
continuum.
(Boulanger et al. \cite{boulanger}).  
Indeed the three PAH features at 
$\lambda$6.2, 7.7, and 8.6 $\mu$m in M~82 can be fit without any
significant underlying continuum at all; a continuum
contribution is fit with a simple linear curve, increasing
with wavelength 
(Laurent et al. \cite{laurent}).  
Vogler et al. (\cite{vogler}) also
find that a continuum contributes
only $\sim$ 5\% for these three PAH bands.  As this is less than
a typical error bar on the observations, we consider the continuum
contribution below these PAH features to be negligible.

Similar arguments apply to the ionic emission lines.  In nuclear
regions,
a [NeII] emission line at
 12.8 $\mu$m has been observed 
 and is
blended with the 12.7 $\mu$m PAH feature.
However, in a sample of 69 nearby normal spiral galaxies with
ISOCAM spectra,
Roussel et al. (\cite{rousselc}) found that the contribution from the
[NeII] line is negligible.  The $\lambda$7.0 $\mu$m [ArII]
line, visible at the nucleus of M~83 (Fig.~\ref{m83_spectrum})
 is no longer seen in the interarm regions of this galaxy (Vogler et al.
\cite{vogler}).

The final possible MIR feature to be remarked upon is a broad silicate
absorption feature at 9 $\mu$m which has been
detected in narrow central regions of very 
dusty starbursts and AGNs (e.g. 
Laurent et al. \cite{laurent}; 
Tran et al. \cite{tran}; Sturm
et al. \cite{sturm00}). 
This feature is not likely to be important
toward the central regions of normal 
starbursts or in quiescent disks
(e.g. 
Roussel et al \cite{roussela}, \cite{rousselb}; 
Vogler et al \cite{vogler},
Dale et al. \cite{dale01}, also arguments in
Spoon et al. \cite{spoon}),
even in the case of edge-on spiral galaxies (Mattila et al.
\cite{mattila99}, for  NGC891). 

Therefore, the MIR spectrum of a quiescent galaxy like NGC~5907 
can be expected to be dominated by the PAH emission bands, with
a weak underlying continuum which becomes stronger at longer
wavelengths.  
We thus interpret the emission in the various observed bands as
follows:

\noindent {\bf 6.8N:}  Lorentzian wings of the $\lambda$6.2 $\mu$m and 
$\lambda$7.7 $\mu$m PAH bands.

\noindent {\bf 7.7N:}  $\lambda$7.7 $\mu$m PAH band.

\noindent{\bf 9.6N:} Lorentzian wings of the
$\lambda$7.7 $\mu$m, $\lambda$8.6 $\mu$m,
and $\lambda$11.3 $\mu$m  PAH bands plus possible minor contribution
from a continuum.

\noindent{\bf 11.3N:}  $\lambda$11.3 $\mu$m PAH band plus possible 
continuum.

\noindent{\bf 6.7W:}   Combined $\lambda$6.2,
$\lambda$7.7, and part of the $\lambda$8.6 $\mu$m PAH band emission.

\noindent{\bf 12W:}  Broad band emission, equivalent to the 
IRAS $\lambda$ 12 $\mu$m band, dominated by PAH features with
likely continuum.

\subsection{The Stellar Contribution}
\label{stellar_contribution}

The stellar contribution to the MIR band is illustrated beautifully
in Lu et al. (\cite{lu}) for both a reddened and de-reddened case (their
Fig. 6).  These authors have obtained complete ISO spectra in
the 2.4 to 5.9 $\mu$m and 5.8 to 11.6 $\mu$m bands for 45 galaxies
finding, as have other authors, that the characteristic shape of
the spectrum in these bands is extremely consistent,
galaxy to galaxy.
The stellar contribution can be approximated by a 
modified black body
%of either 750 K or 1000 K temperature (depending on reddening)
with a $\lambda^{-2}$ emissivity law.  This behaviour for
the stellar contribution also accurately describes the elliptical
galaxies in their sample.  Thus,
we expect the stellar contribution
in both the disk and halo regions 
of NGC~5907 to have a similar $\lambda$ dependence.

We have obtained the K$_s$ band ($\lambda$ 2.159 $\mu$m)
 image for NGC~5907, from 
the 2MASS Large Galaxy Atlas (LGA, 
Jarrett et al. \cite{jarrett}),
 which is the band most likely to
show only stars with negligible dust obscuration.  This image was
interpolated onto the same grid as our MIR images and then smoothed
to 7.2$^{\prime\prime}$ resolution.  The `data numbers' ($dn$), in
the FITS file were converted to Jy using the transformation,
$f_\nu\,=\,6.248\,\times\,10^{-6}\,dn$ 
Jy \footnote{This transformation uses 
the calibration, $m_{K_s}\,=\,ZP\,-2.5\,log(dn)$ and
 $f_\nu\,=\,f_{\nu}(0)\,10^{-0.4\,m_{K_s}}$
 where $ZP$ = 20.0704994 is the zero point
provided in the FITS header, and $f_{\nu}(0)\,=\,666.7$ Jy, the latter
from
{\it http://ssc.spitzer.caltech.edu/documents/cookbook/html/cookbook.html.}}.
As a check on our processing steps, 
we measured the total flux density of the source
in the regridded, smoothed image, finding
$f_\nu\,=\,1.322$ Jy which corresponds to 
$m_{K_s}\,=\,6.76$.  This agrees with the magnitude
given in the LGA for NGC~5907 of $m_{K_s}\,=\,6.76\,\pm\,0.02$.
The resulting image has an
rms noise of $\sigma\,=\,1.6$ $\mu$Jy arcsec$^{-2}$.
% is shown
%as a greyscale to enhance the lowest levels of emission, with the
%contours of Fig.~\ref{lw6} superimposed,  
%in Fig.\ref{stellar_halo}.  
From the emissivity law (following Lu et al.),
we expect the stellar contribution,
extrapolated into the MIR bands in which PAHs dominate to be
7.9\% in the 7.7N band,
3.7\% at 11.3N,
and 10\% at 6.7W 
of whatever value is in the K$_s$ band at the relevant location.
With this extrapolation, we created 3 maps from the K$_s$ band image
for the 3 bands dominated by PAH emission and subtracted each from
the total emission map.  These results are shown in the insets to
Figs.~\ref{lw6}, \ref{lw8}, and \ref{lw2} with the same contour levels
as the main maps.  We then measured the global flux of the
stellar-subtracted maps and compared this to the global flux of
the main maps finding a difference of 5\%, 2\%, and 2\% for
the 7.7 $\mu$m, 11 $\mu$m, and 6.7 $\mu$m  maps, respectively
(ignoring emission from the two outer frames in the latter case).
Since the flux corrections are well within other error bars
(Sect.~\ref{observations_reductions}) and the 
appearance of the maps is also  close to the original non-subtracted
maps, we consider the stellar contribution to be negligible and continue
our discussion and analysis with respect to the unsubtracted maps.

%______________________________________________________________

\section{Results}
\label{results}

All maps along with the best estimated error maps are shown in
Figs.~\ref{lw5} to \ref{lw2} as described in Sect.~\ref{reductions}.
The error maps 
delineate the final field of view.
Emission has been detected in every observing band as shown in 
these figures.

\subsection{Disk Emission}
\label{disk_emission}

Emission is clearly detected along the disk of NGC~5907.
In no individual band do we have full coverage of the entire galaxy.
However, for the narrow bands, 
6.8N, 7.7N, 9.6N, and 11.3N,
 we have
full coverage of the south-east major axis and for 6.7W, small
segments at both ends of the galaxy have been observed.  In all
cases, the emission extends along the optical disk with detections
to the `end' of the optical disk in some cases.  

The extent of the PAH emission, specifically, is
 illustrated by
Fig.~\ref{lw6_dss}a in which we show the smoothed 7.7N emission
($\lambda$ 7.7 $\mu$m PAHs)
over the DSS optical image.  It is clear from this overlay that
the PAH emission extends all along the optical disk
of the galaxy.  
It is further illustrated in Fig.~\ref{lw2_mom0}
in which we show the smoothed, high sensitivity
6.7W emission ($\lambda$6.2, 7.7, and 8.6 $\mu$m PAHs) 
superimposed on the HI total intensity map from
Shang et al. (\cite{shang}).  
 PAH emission
is seen very far out along the galaxy disk where the HI emission
is also strong.  The total extent of the PAH emission along the
major axis is 11.0 arcmin ($\sim$ 35 kpc), as measured to the outermost
detectable contiguous features on the unsmoothed map
(cf. 41 kpc for the optical disk, Table~\ref{basic_parameters}).  
However, there is also
some evidence for real features farther out.  For example, several
emission peaks can be seen at the edges of the 
6.7W field of view on
the south-east end of the major axis at
RA = 15$^{\rm h}$ 16$^{\rm m}$ 12$^{\rm s}$, DEC =
56$^\circ$ 13$^{\prime}$ 09$^{\prime\prime}$. A channel-by-channel
comparison
(in velocity) of the HI distribution with the PAH emission, however, fails to
show any clear correlation.

\begin{figure}%[h]
%\resizebox{\hsize}{!}{\includegraphics{fig8.ps}}
\caption{{\bf (a)} The 7.7N band image
($\lambda$7.7 $\mu$m PAH band), 
smoothed to 12$^{\prime\prime}$ resolution
superimposed on the optical DSS image.
Contours are at
4.6 (2$\sigma_{map}$), 7.5, 13, 25, 50, 75, 200, and 380
$\times$ 10$^{-3}$ mJy arcsec$^{-2}$.
{\bf (b)} The 7.7N band image as in (a) superimposed on
a greyscale 
$\lambda$850 SCUBA $\mu$m map from Alton et al. (\cite{alton04}) with a resolution
of 16$^{\prime\prime}$.  Note that the SCUBA field
of view ends at a Declination of 56$^\circ$ 17$^\prime$.}
\label{lw6_dss}
\end{figure}

\begin{figure}%[h]
%\resizebox{\hsize}{!}{\includegraphics{fig9.ps}}
\caption{The 6.7W band image, smoothed to 12$^{\prime\prime}$ resolution
superimposed on the HI total intensity image (data from Shang et al.
\cite{shang}).
Contours are at
0.0004 (2$\sigma_{map}$), 0.0008, 0.0015, 0.003, 0.01, 0.04, 
and 0.10 mJy arcsec$^{-2}$.} 
\label{lw2_mom0}
\end{figure}

The lack of {\it global} correlation with HI is further
illustrated in Fig.~\ref{major_axis}a in which we show
major axis slices of the HI total intensity distribution, the
CO, and the smoothed
7.7N emission.  It is clear from this plot that
the PAH emission follows the molecular gas distribution
and not the atomic gas distribution.  
This was also found by
Dumke et al. (\cite{dumke97}) 
for the $\lambda$1.2 mm dust distribution
as well as by Alton et al. (\cite{alton04}) for the  $\lambda$850 $\mu$m emission. 
Fig.~\ref{major_axis}a  shows two sets of peaks 
in the $\lambda$7.7 $\mu$m distribution on either side
of the nucleus, one at $\,\sim\,\pm\,70^{\prime\prime}$ and the second
at $\,\sim\,\pm\,100^{\prime\prime}$. 
The CO comparison also show two peaks, interpreted by Dumke et al. 
(\cite{dumke97})
to represent rings or spiral arms, but at somewhat offset
  radii of $\pm\,60^{\prime\prime}$ and
$\pm\,120^{\prime\prime}$.  

\begin{figure}[h]
%\centerline{\includegraphics*[scale=0.6]{majoraxis_cuts.ps}}
%\resizebox{\hsize}{!}{\includegraphics{fig10.ps}}
\caption{Total intensity cuts along the major axis of NGC~5907.
 Positional 
uncertainties are of order
$\pm$7$^{\prime\prime}$.
{\bf (a)} The HI total intensity (faint extended
curve) from data of Shang et al. (\cite{shang}), the CO distribution 
(faint solid plus dashed curves showing data points and error bars)
from Dumke et al. (\cite{dumke97}), and the $\lambda$7.7 $\mu$m from
this work (darkest curve), the latter smoothed to the same resolution
as the HI data 
(18.6$^{\prime\prime}$$\,\times\,$17.8$^{\prime\prime}$ at PA =
-38$^\circ$).  The CO resolution is 21$^{\prime\prime}$ and the
center point has been placed 
approximately at the CO center of RA(J2000) = 
15$^{\rm h}$ 15$^{\rm m}$ 53.8$^{\rm s}$, 
DEC(J2000) = 56$^\circ$ 19$^\prime$ 37$^{\prime\prime}$.
{\bf (b)}  The $\lambda$7.7 $\mu$m PAH emission from the 
7.7N band,
smoothed to 16$^{\prime\prime}$ resolution shown with
the $\lambda$850 $\mu$m SCUBA emission at the same
resolution.  Curves and intensities represent an average over a 
16$^{\prime\prime}$ swath parallel to the major axis. The arrow
marks the edge of the SCUBA field of view.}
\label{major_axis}
\end{figure}

In Fig.~\ref{major_axis}b, we show a comparison
between major axis cuts for our ISO $\lambda$7.7 $\mu$m data and the  
$\lambda$850 $\mu$m SCUBA data, both averaged over
a minor axis extent of 16$^{\prime\prime}$ and at the same spatial resolution.
  The shape of the PAH emission curve and the 
$\lambda$850 $\mu$m emission curve, the latter which
traces cool dust, are remarkably similar showing
the same number and approximate, but not exact, positions of
 peaks on either side of the nucleus. Positional offsets between
between the two different slices
for 5 peaks range from zero to 19$^{\prime\prime}$
with a typical value of $\sim$ 13$^{\prime\prime}$.  Given that
the resolution is 16$^{\prime\prime}$ and relative positional uncertainties
between maps is of order 7$^{\prime\prime}$ (see caption), the
positions of the peaks are likely consistent, though we cannot
rule out the possibility of some displacement which might be
detected with higher resolution observations.
The
line ratios from these curves (calculated though not plotted)
 are also consistent with the values of Haas et al. (\cite{haas}) who
found $\lambda$7.7 $\mu$m/$\lambda$850 $\mu$m = 2 
(1.8 to 2.2 for 5 galaxies).  Their denominator is from SCUBA observations
but the numerator represents the peak  $\lambda$7.7 $\mu$m flux rather than
the average in the 
7.7N band as we have plotted.  All of the emission in our
7.7N band is from the $\lambda$7.7 $\mu$m PAH line but the peak, if we
could measure it directly, would
be higher than the average.  From Fig.~\ref{spectrum}, the increase should
be of order a factor of $\approx$ 1.5 which would put our ratio in even
closer agreement with Haas et al.  Therefore, overall, our results 
are consistent with a possible
relationship between PAHs and the 850 $\mu$m cool dust distribution.

Finally, there is excess $\lambda$7.7 $\mu$m emission
between radii of -150$^{\prime\prime}$ and
-250$^{\prime\prime}$ (Fig.~\ref{major_axis}a) in comparison to
the CO distribution.  (Note that we cannot discern this in 
Fig.~\ref{major_axis}b because of the truncated field of view of
the SCUBA map.)  We believe that this excess is significant.  Indeed,
Dumke et al. (\cite{dumke97}) 
also found an excess in the $\lambda$1.2 mm
emission in comparison to CO over the same radii but on both the north
and south sides of the nucleus.
This shows that there is 
PAH emission as well as cool dust
where there is little molecular gas.  We will return to
this point in Sect.~\ref{discussion}.

In Fig.~\ref{lw10_dss} we show the
broad band 12W emission
(equivalent to the IRAS 12 $\mu$m band)
superimposed on the DSS optical image.
This MIR band should contain the highest contribution of
continuum emission from VSGs
 compared to the other 
bands (Fig.~\ref{m83_spectrum}, inset).  The 12W band
has the highest spatial resolution of all observations (1$^{\prime\prime}$)
and reveals 
that the strongest ridge of MIR emission lies along the
optical dust lane (note that the field of view does not cover the entire
disk), indicating a good correlation between VSGs and the optically
obscuring dust.

\begin{figure}[h]
%\epsscale{.80}
%\centerline{\includegraphics*[scale=1.0]{lw10_dss.ps}}
%\resizebox{\hsize}{!}{\includegraphics{fig11.ps}}
\caption{The IRAS band-equivalent image, 12W,
(contours) over the Digitized Sky Survey (DSS) optical image.  
Contours are at
0.028, 0.050, 0.10, 0.18, 0.28, and 0.36 mJy arcsec$^{-2}$.  The
12W field of view does not extend over the whole optical image
(see Fig.~\ref{lw10}b). 
The known faint stellar ring (Sect.~\ref{ngc5907}), which cannot
be discerned in this image,
 intersects the disk
at declinations of 56$^\circ$ 23.4$^\prime$
and 56$^\circ$ 16.8$^\prime$
in projection (Zheng et al. \cite{zheng}).
\label{lw10_dss}}
\end{figure}

\subsection{High Latitude Emission}
\label{high_latitude_emission}

A particularly striking result
is the evidence for high latitude emission,
especially in the 7.7N band
(Fig.~\ref{lw6}), the
 11.3N band  
(Fig.~\ref{lw8}),
and the  
high-sensitivity beam-switched 6.7W band 
(Fig.~\ref{lw2})
which all trace  PAH emission.
The  11.3N band  
in principle, can contain PAHs plus continuum, but 
the continnuum should be negligible so far from star forming regions. Thus it 
is clear that PAHs exist far from the plane
of NGC~5907, in the case of 6.7W,
as far as 6.5 kpc (2$^{\prime}$)
from the plane, though $\sim$ 3 kpc is more typical.  Even the 6.8N
band (Fig.~\ref{lw5}), which should
consist only of emission from PAH band wings, 
shows some evidence for features
away from the plane.  
Given the galaxy's edge-on orientation 
(86.5$^\circ$, Table~\ref{basic_parameters}), the correction for
galaxy inclination is negligible. For example, for a PAH disk
radius of 250$^{\prime\prime}$ (Fig.~\ref{major_axis}), the semi-minor
axis is only 15.3$^{\prime\prime}$ in a thin disk model.
Moreover, the CO semi-minor axis is only $\sim$ 4 arcsec if we
use the HWHM of 200 pc modeled by Dumke et al. (\cite{dumke97}).
These features resemble the extra-planar emission seen in other galaxies, 
 (see, e.g. the radio continuum features in NGC~5775,
Lee et al. \cite{lee01})
and some are similarly arc-like in appearance (e.g. the east side of
the $\lambda$7.7 $\mu$m emission in Fig.~\ref{lw6}).  Recall also
(Sect.~\ref{ngc5907})
that the stellar scale height has been measured to be 0.49 kpc
(9$^{\prime\prime}$).  Note that the narrow bands, 7.7N  and 
11.3N, show more structure than the wide band, 6.7W  which may
result from the ability of these bands to isolate a single PAH
line.  Although structural details may vary, both narrow bands
show  
high latitude features at approximately the same locations
(see also Sect.~\ref{errors}).  

Given that the beam-switched 
6.7W band observations are our most sensitive
and also the fact that this band is dominated by several PAH emission
features, we have taken the original, unsmoothed 
6.7W data shown in
the central regions of Fig.~\ref{lw2} and averaged the emission in
strips parallel to the major axis.  The resulting
minor axis profile is
shown in Fig.~\ref{profile}. The wings on this averaged
profile approach the noise between 100$^{\prime\prime}$ and
150$^{\prime\prime}$ (5.3 kpc to 8.0 kpc)
and the line half-width at the 3$\sigma$ level is
$\sim$ 95$^{\prime\prime}$ (5 kpc).
We fit various functional forms to the profile up to 3 components
in total, including Gaussians, Lorentzians, Voigt, 
combinations
of these forms and a combination of Gaussian $+$ exponential.
The best fit results were achieved by
3 Gaussians or 1 Gaussian plus 1 Voigt profile, which resulted
in equivalent residual rms values.  However, since the Voigt profile
does not have easily identifiable parameters, we show the fits 
(Fig.~\ref{profile}) 
for the Gaussian combinations only
and provide tabular information
(Table~\ref{profile_table}) 
for these combinations 
as well as the Gaussian $+$ exponential for
comparative purposes.  The broadest 
Gaussian scale length is 65.3$^{\prime\prime}$ (3.5 kpc).
Increasing the number of components
would reduce the residual rms further but such a situation
is likely not physical.  The profile, in general, consists of
a central Gaussian component plus broad wings.  
 Also, since the high latitude emission is highly structured
(Fig.~\ref{lw2}), the scale length of the broad component
will vary with position.
  Nevertheless, it
is clear that an extended halo or thick disk of PAH emission
is present in NGC~5907 with a characteristic scale height of
3.5 to 5 kpc.

\begin{table*}
\caption{6.7W z Profile\label{profile_table}}
\centering
\begin{tabular}{lcccc}
\hline\hline
{Data or Type of Fit} & Component No. &
{Peak Intensity\hfill} & {FWHM}$^\mathrm{a}$ & 
{Residual rms}  \\
  & & (mJy arcsec$^{-2}$)  & (arcsec) & (mJy arcsec$^{-2}$)\\
\hline
Data                 &   &       &       & 0.20 $\times$ 10$^{-3}$ \\
1-component Gaussian & 1 & 0.315 & 18.3  & 3.96 $\times$ 10$^{-3}$ \\
2-component Gaussian & 1 & 0.287 & 16.2  &  \\
            & 2 & 0.0376 & 43.2  & 1.07 $\times$ 10$^{-3}$ \\
Gaussian $+$ Exponential & 1 & 0.264 & 16.6 &  \\
  & 2 & 0.065 & 16.5$^\mathrm{b}$ & 1.11 $\times$ 10$^{-3}$ \\
3-component Gaussian & 1 & 0.257 & 15.5  &  \\
            & 2 & 0.0564 & 27.2  & \\
            & 3 & 0.00131 & 65.3  & 0.86 $\times$ 10$^{-3}$ \\
\hline
\end{tabular}
\begin{list}{}{}
\item[$^{\mathrm{a}}$]
Unless otherwise indicated.
\item[$^{\mathrm{b}}$]
Exponential scale length.
\end{list}
\end{table*}

\begin{figure}[h]
%\epsscale{.80}
%\centerline{\includegraphics*[scale=1.0]{profile.ps}}
%\resizebox{\hsize}{!}{\includegraphics{fig12.ps}}
\caption{Average profile of the PAH emission as a function of z
from the high sensitivity 6.7W data (Fig.~\ref{lw2}).  Negative
values of z represent the western side of the major axis. {\bf (a)}
Plot of the data over the full range of intensity.
{\bf (b)}  Plot of the data and two of the models over only the
lowest 1.7\% of the total intensity.  The dark solid curve shows
the data with the noise, at an rms level of $2\,\times\,10^{-4}$
mJy arcsec$^{-2}$, now obvious.  
The grey curve shows the best fit 2-component gaussian
model and the narrower dark solid curves shows the best fit 3-component
gaussian model.  See Table~\ref{profile_table} for the parameters of
the gaussian fits.}
\label{profile}
\end{figure}

Given the apparent correlation between 
the ISO $\lambda$7.7 $\mu$m emission and the
SCUBA $\lambda$850 $\mu$m emission in the disk (Sect.~\ref{disk_emission}),
it is of interest to see if this correlation extends to the halo region.
Fig.~\ref{lw6_dss}b shows a comparison between these bands.
Note that the $\lambda$850 $\mu$m field of view is truncated at a Declination
of 56$^\circ$ 17$^\prime$ and the 
7.7N map is truncated on the
eastern side.  Note also that the greyscale showing apparent emission away
from the plane in the SCUBA data is all below the 3$\sigma$ level
(Alton et al. \cite{alton04}, 
their Fig. 1) with the exception of the 'detached' feature
at RA = 15 15 52, DEC = 56 17 45.
%There are hints of some correlation, for example at
%RA = 15 15 50, DEC = 56 18 15 (an extension towards the west)
%RA = 15 15 45, DEC = 56 29 45 (a broad extension towards the west)
%RA = 15 16 00, DEC = 56 19 15 (an extension towards the east) and
%RA = 15 15 55, DEC = 56 21 00 (an extension towards the east).
%However, given the low signal-to-noise of the SCUBA features, we
%do not consider this to be definitive.  
Indeed, Alton et al. find
an exponential scaleheight of only 0.11 kpc 
(2$^{\prime\prime}$) for the $\lambda$850 $\mu$m
map.  If the $\lambda$850 $\mu$m emission
existed in the halo at the same fraction of the in-disk emission
as the PAHs, then the $\lambda$850 $\mu$m emission should
have been detected in the halo above the noise
of the Alton et al. map.  This suggests
that either the large grains are under-represented in the halo 
in comparison to the PAHs, or
are under-emitting.  

The broad band 
12W emission shows very little vertically
extended emission
(Fig.~\ref{lw10},
Fig.~\ref{lw10_dss})
 in comparison to the PAH bands
(cf. Figs.~\ref{lw6} and \ref{lw8}).
This is likely due to 
the fact that the dynamic range in the 12W band is
only 10 to 17\% as great as in the 7.7N and 11.3N bands
(Table~\ref{observing_map}).
For example, the extended vertical emission seen in Figs.~\ref{lw6}
and \ref{lw8}, {\it if} present in Fig.~\ref{lw10}
at the same relative brightness
with respect to the maximum disk emission, would mostly fall at
the level of the noise or lower.
  One obvious feature is visible at the
edge of the field of view on the north, extending to the east.
(The optical feature within this extension appears to be a foreground
star.)  The reality of the large scale feature is not clear but
its `footprints' close to the disk have  
counterparts in the PAH bands 
(e.g. Fig.~\ref{lw8}) suggesting that the PAH contribution within
this band may be responsible for it.

%This image is shown superimposed on the a greyscale
%of the LW8 $\lambda$7.7 $\mu$m PAH emission, over the same
%field of view and smoothed to the same resolution, in Fig.~\ref{lw8_lw10}.
%This image shows the similarities and contrast between the two bands.
%The LW8 band, which contains PAH emission alone, shows some
%correlation with the broad LW10 band, which contains both PAH emission
%and continuum.  However, the PAH band alone shows
%much more emission on the east side which is clearly halo emission. 
%On the east side, LW10 is more extended

\subsection{Flux Densities and Band Ratios}
\label{ratios}

In no case do we have complete coverage of
the galaxy so global flux densities cannot be obtained.  However,
we have determined the flux density for each of the fields displayed.
These are listed in Table~\ref{fluxes}.  Since 
the 6.8N through 11.3N bands have
the same field of view, their fluxes can be compared directly.  There
is one common field for all maps, however, which is outlined in
Fig.~\ref{lw2}b.  Thus, we also calculate fluxes for this common
region (Table~\ref{fluxes}) and compare these results for all bands.
The resulting flux ratios are
 provided in Table~\ref{flux_ratios}
with some comparative values, where known, also given.

\begin{table*}
\caption{Flux Densities\label{fluxes}}
\centering
\begin{tabular}{lcc}
\hline\hline
{Band\hfill} & {Flux Density (Field)$^\mathrm{a}$} & 
{Flux
Density (Common)$^\mathrm{b}$}   \\
  & (mJy) & (mJy) \\
\hline
6.8N  & 1170 $\pm$ 97  & 519 $\pm$ 42 \\
7.7N  & 2170 $\pm$ 141 & 918 $\pm$ 60 \\
9.6N & 919$^\mathrm{c}$ $\pm$ 66 & 360 $\pm$ 26 \\
11.3N & 2071 $\pm$ 116 & 828 $\pm$ 46 \\
12W & 940 $\pm$ 38 & 557 $\pm$ 57 \\
6.7W & 666 $\pm$ 22 & 591 $\pm$ 20 \\
\hline
\end{tabular}
\begin{list}{}{}
\item[$^{\mathrm{a}}$]
 Flux density determined over the entire field shown in
Fig.~\ref{lw5}a through Fig.~\ref{lw2}a.
The error (here and throughout) 
is calculated 
via $\sqrt{\Sigma \sigma_i^2}$ over the map pixels, $i$, 
where $\sigma_i$ is the quadratic sum at each pixel of
$\sigma_{rms}$, the
Transient rms, the $1\sigma$ sky error {\it and} the calibration error (see
Table~\ref{observing_map}).
%in practice, to do this, take the err^2 map times 36 + (calerr*flux)^2
%then sqrt
%e.g. lw5:  err^2map: mean=3.1172e-4 npix=153983 so total=47.99958
%then correct for npix smoothing to get total = 47.99958*36=1.72798e3
%then the cal error which is 7.5%, i.e. (7.5/100*1170)^2=7.7e3
%then add the errors= 1.72798e3+7.7e3=9.428e3
%finally take the sqrt =97.098 and that's the error on the flux of 1170 
%for lw10 don't multiply by 36 because there was no smoothing
\item[$^{\mathrm{b}}$]
Flux density determined over the region that is common to
all maps and outlined in Fig.~\ref{lw2}b. 
\item[$^{\mathrm{c}}$]
Excluding the 3 small regions with residual errors (see
Sect.~\ref{errors}).
\end{list}
\end{table*}

The 6.7W:12W ratio is interesting since it is
significantly higher than the values given
by Dale et al. (\cite{dale01}).  This ratio
represents a comparison between a broad ISOCAM band that 
contains mainly 3 PAH bands (Sect.~\ref{spectrum}) 
with the broader 12 $\mu$m IRAS-equivalent band
which may contain both PAHs and continuum.  
Dale et al. (\cite{dale01}) 
determined average values of this ratio,
using the IRAS 12$\mu$m flux itself,
 for normal galaxies in different
$f_{60}/f_{100}$ bins. 
The IRAS flux ratio, $f_{60}/f_{100}$
is a measure of the heating of classical grains in temperature
equilibrium.  As such, it is widely used
as a measure of star formation activity, with 
an increasing ratio implying greater dust heating
(Helou \cite{helou86}; Dale et al. \cite{dale00}, \cite{dale01}). 
As the SFR increases ($f_{60}/f_{100}$ increases),
the 6.7W/(broad-band 12$\mu$m) ratio decreases, reflecting
the fact that a) the contribution
of a hot dust continuum increases with SFR
and b) PAHs are
 expected to be destroyed
in regions of high heating intensity such as active star forming
regions (Dale et al. \cite{dale01} and references therein). 
  The 
$f_{60}/f_{100}$ ratio for NGC~5907
(0.24, Table~\ref{basic_parameters}) places it 
at a SFR that falls below all of the galaxies of the
Dale et al. sample.
Moreover,
the $f_{60}/f_{100}$ ratio
for NGC~5907 could be even lower than that quoted in
Table~\ref{basic_parameters}, i.e. 0.21
from earlier ISO data, Bendo et al. \cite{bendo02}.
Thus, the high 
6.7W:12W ratio for NGC~5907
(Table~\ref{flux_ratios}) is likely consistent with the data
from other galaxies in that it would 
represent an extension to
lower SFR. 

\begin{table*}
\caption{Flux Density Ratios\label{flux_ratios}}
\centering
\begin{tabular}{lccc}
\hline\hline
{Bands$^\mathrm{a}$\hfill} & {Ratio or Relation} & 
{Galaxy or Region} &{Ref.}  \\
\hline
6.7W:12(IRAS) & [0.52$^{+0.22}_{-0.15}$]:1$^\mathrm{b}$ 
& 
8 galaxies & Dale et al. \cite{dale01} \\
6.7W:12W & [1.06 $\pm$ 0.16]$^\mathrm{c}$:1  
& Common$^\mathrm{d}$ 
& This Work \\
6.8N:7.7N:9.6N:11.3N & 
[0.57 $\pm$ 0.08]:1:[0.39 $\pm$ 0.05]:[0.90 $\pm$ 0.11]
& Common$^\mathrm{d}$ & This Work \\
6.8N:7.7N:9.6N:11.3N & 
[0.54 $\pm$ 0.08]:1:[0.42 $\pm$ 0.06]:[0.95 $\pm$ 0.12]
& Field$^\mathrm{d}$ & This Work \\
6.8N:7.7N:9.6N:11.3N & 
0.49 : 1 : 0.59 : 1.12$^\mathrm{e}$
& M~83 (R $\le$ 1$^\prime$)$^\mathrm{f}$ &
Vogler et al. (\cite{vogler}) \\
6.8N:7.7N:9.6N:11.3N & 
0.55 : 1 : 0.48 : 0.95$^\mathrm{e}$
& M~83 (1$^\prime$ $\le$ R $\le$ 3$^\prime$)$^\mathrm{f}$ 
&
Vogler et al. (\cite{vogler}) \\
\hline
\end{tabular}
\begin{list}{}{}
\item[$^{\mathrm{a}}$]
The center wavelengths of the bands is given below each
sequence that is the same.
\item[$^{\mathrm{b}}$]
Average of 8 galaxies in a 
flux ratio, $f_{60}/f_{100}$, bin
from 0.28 to 0.35 (Dale et al. \cite{dale01}).
\item[$^{\mathrm{c}}$]
This error includes an additional 5\% error to account for
variations in orbit (Sect.~\ref{errors}).
\item[$^{\mathrm{d}}$]
Common or Field region as quoted in Table~\ref{fluxes}.
\item[$^{\mathrm{e}}$]
Errors (estimates only) are believed to be of order 20\%.
\item[$^{\mathrm{f}}$]
R is the projected radial distance from the center of the galaxy.
\end{list}
\end{table*}

The ratios for the bands 6.8N through 11.3N
(Table~\ref{flux_ratios}) 
 all have the same field of view and so can be compared directly.  
(Note also, that all results for the 'field' and for the 'common'
regions agree within errors.)
For these bands, we
make a comparison to the galaxy, M~83, for which similar observations
have been made.  NGC~5907 is not perfectly identical to M~83 in its
global properties (Table ~\ref{basic_parameters}).  The two galaxies
are of similar morphological types and have
similar global
star formation rates, infrared luminosities, and modelled
temperatures.
  However, 
M~83 is physically smaller than NGC~5907.
Thus, NGC~5907
contains an order of magnitude more dust but 
M~83 is the more
active galaxy when SFR per unit area is determined.
Nevertheless,  the shapes of the different spectra in
M~83 (see Vogler et al. \cite{vogler} and Fig.~\ref{m83_spectrum})
are remarkably constant as a function of location
in that galaxy, i.e. as a function of differing local SFR,
the main differences being increased contributions of
continuum, especially 
longwards of 
14$\mu$m (which we sample only marginally in NGC~5907 in the 
12W band).
For M~83, the main differences between spectra can be seen as
an increased continuum contribution from
 hot grains which become
more prominent towards the nucleus.  
Thus, we provide the 
6.8N through 11.3N 
 ratios for M~83 for the
inner, hotter, 1$^{\prime}$ radius region as well as the outer, cooler,
region between radii of 1$^{\prime}$ and 3$^{\prime}$ in
Table~\ref{flux_ratios}.
Given the estimated errors on the various ratios, there is agreement
between all values for M~83 and those of NGC~5907, although
the most variation is seen for 9.6N. 
 The band, 9.6N, measures a mixture
of PAH wings plus possible hot continuum. 
The 9.6N:7.7N ratio for
NGC~5907 is  somewhat low in comparison to the nucleus of M~83 but agrees
with the result for the interarm region of M~83.  This is
again quite consistent with the low SFR for NGC~5907.

The Dale et al. comparison values for 
6.7W:12W are globally
determined values.  Our data allow us to further
examine this ratio as a function
of projected distance from the major axis, as shown in Fig.~\ref{zratio}.
In  Fig.~\ref{zratio}a, we show the increasing
6.7W:12W ratio with projected distance from
the major axis.  Given that this ratio can only be determined over
a restricted area of the 'common' 
region due to signal-to-noise considerations, 
this may reflect either
an increase in this ratio with {\it z} height or an increase with
distance from the center of the galaxy
since the region shown does not extend beyond the projected
optical disk.  In either case, this
is likely due to an increase in the distance from star forming
regions and therefore a decrease in the contribution of the
hot continuum that would be expected in the 12W band.
In Fig.~\ref{zratio}b, we show
the ratio of 
the emission in two PAH bands: 11.3N/7.7N. 
  In this case, the
ratio can be determined to much higher latitude.  
The maximum
change in ratio is of order 2.4 with the in-disk value lower.
Given that more continuum, if present,
 is expected in the 11.3N band than
the 7.7N band in the disk, then the in-disk ratio could be
lower than shown.
As indicated above,
 the global ratio of these two bands is quite consistent
with what is seen in other galaxies so we may be observing
a real variation in the 11.3/7.7 $\mu$m PAH ratio between
the changing environments of the disk and halo.  The
nature of this variation is unclear and the exact 
magnitude will require the acquisition of spectra and
modelling.  However, we note that the variation seen here
is not inconsistent with the variations between PAH feature
strengths observed for regions in our own Milky Way, i.e.
of order a factor of 5 (see Peeters et al. \cite{peeters02}).

\begin{figure}%[h]
%\epsscale{.80}
%\centerline{\includegraphics*[scale=0.5]{zratios.ps}}
%\resizebox{\hsize}{!}{\includegraphics{fig13.ps}}
\caption{{\bf (a)}  Profile of the 
 6.7W/12W ratio from the common
region only (see Fig.~\ref{lw2}b)
as a function of projected distance from the major
axis.  The values are averages over
71$^{\prime\prime}$ strips parallel to the major axis.  
{\bf (b)} The {\it z} profile of the 11.3N/7.7N ratio for the
fields shown in Figs.~\ref{lw6} and \ref{lw8}, but averaged
over 320$^{\prime\prime}$ strips.  Note that the {\it z} extent
is much larger than in (a).
%(A 4$^{\prime\prime}$
%shift in RA was first applied to LW2 to align the major axes.)
\label{zratio}}
\end{figure}

\subsection{Radio Continuum Correlation?}
\label{radio_correlation}

Vogler et al. (\cite{vogler}) find a good spatial correlation between
the $\lambda$6 cm radio continuum emission and the 
6.7W (largely PAH)
MIR emission in M~83, and 
Brar et al. (\cite{brar03}) also find a good
correlation between the $\lambda$850 $\mu$m emission and
the 617 MHz radio continuum emission in the edge-on
galaxy, NGC~5775.  These results, which link the MIR or sub-mm with
the radio continuum, are reminiscent of the 
well-known FIR-radio correlation whose origin is not
fully understood but is thought to
relate to star formation in galaxies (see 
Groves et al. \cite{groves}
and references therein).  Thus,
a comparison of our ISO results with radio data is of
interest.  
The best available radio data sets are those of the
NRAO VLA Sky Survey (NVSS) at 1420 MHz 
(Condon et al. \cite{condon})
and the Westerbork Northern Sky Survey
(WENSS) at 327 MHz 
(Rengelink et al. \cite{rengelink}).
The radio emission in the 
in the VLA
Faint Images of the Radio Sky at Twenty-Centimeters
survey (FIRST) is
 too faint for comparison (but see note below).  In
Fig.~\ref{lw6_radio}, we show the 
7.7N ($\lambda$7.7 $\mu$m PAH)
emission, which is unlikely to contain any significant continuum
component,
in comparison to these two radio images.  
The strong radio
source at the end of the south-eastern major axis
(actually a double) is believed to be from a background
 source (Dumke et al. \cite{dumke00}).
There are significant differences between the fields of view
and resolutions of the ISO and radio data sets which
make comparison difficult, and smoothing the 
7.7N emission
results in confusion and 
 truncations near the field edges.  
In addition, details in the structure of the 7.7N map at
low emission levels may be approximate.
Thus any apparent
correlations are tentative and are presented so that higher
resolution data in the future might be compared to these
results.
In particular, a radio continuum extension 
on the west side of the major axis at 
RA $\approx$ 15${\rm h}$ 15${\rm m}$ 45${\rm s}$,
DEC $\approx$ 56$^\circ$ 20$^\prime$
shows extended PAH emission (though the ISO field of view
also truncates here).  South of this feature is another
at
RA $\approx$ 15${\rm h}$ 15${\rm m}$ 50${\rm s}$
DEC $\approx$ 56$^\circ$ 18$^\prime$.
On the east side of the major axis, there is a feature at
RA $\approx$ 15${\rm h}$ 16${\rm m}$ 00${\rm s}$
DEC $\approx$ 56$^\circ$ 20$^\prime$.  These features
in the radio are more easily seen in the 327 MHz image
and at least the northernmost extension on the west side
can also be seen in a 20 cm image of 
Dumke et al. (\cite{dumke00}).
We are not able to determine correlations along the disk
due to the limited ISO field of view and resolution
and also because of confusion with the strong background
source.

\begin{figure*}%[h]
%\includegraphics[width=0.9\textwidth]{lw6_radio.ps}
%\resizebox{\hsize}{!}{\includegraphics{fig14.ps}}
\caption{
{\bf (a)} The 7.7N band image, with contours and field
of view as shown in Fig.~\ref{lw6},
superimposed on the NVSS $\lambda$ 20 cm radio continuum image 
 with low level emission enhanced.
The greyscale ranges from 0.01 ($\approx$ 1$\sigma$)
 to 4 mJy beam$^{-1}$, with a
beam 45$^{\prime\prime}$ in size. The strong
source at the south-east end of the major axis is believed
to be a background source (Dumke et al. \cite{dumke00}).
{\bf (b)} WENSS 327 MHz image (contours)
superimposed on a greyscale 7.7N band image.  Contours
are at 
4.5 (1.5$\sigma$),           6,             9,            12,
15,            20,           30,          40,
70,           110,          and 160 mJy beam$^{-1}$
with a beam size of 54$^{^{\prime\prime}}$ in Right Ascension
and 64.9$^{^{\prime\prime}}$ in Declination.  The greyscale
ranges from 0.003 to 0.02 mJy arcsec$^{-2}$.
}
\label{lw6_radio}
\end{figure*}

Finally, we note that there is a
previously unreported
point source in the FIRST data with
coordinates, 
RA(J2000) = 15 15 49.10, 
DEC(J2000) = 56 20 27.1, which is displaced 
38$^{\prime\prime}$ to the west and 
43$^{\prime\prime}$ to the north of the nucleus
(2.0 and 2.3 kpc, respectively, in projection) with
a flux density of
3.1 mJy.  This flux density corresponds to
a spectral power of 4.5 $\times$ 10$^{19}$
Watts/Hz which is intermediate between that of Cas A
and the brightest supernova remnant in M~82 
(see Irwin et al. \cite{irwin00}) and
is therefore likely a supernova remnant or remnants.

%______________________________________________________________

\section{Discussion}
\label{discussion}

The mid-IR emission in NGC~5907 is consistent with NGC~5907 as a low SFR
galaxy.  As described in Sect.~\ref{ratios}, this galaxy is 
'PAH-rich' and 'hot-dust-poor'.  Our flux ratios are consistent
with those measured in other galaxies at the low end of the
range of SFR and are also 
consistent with those of 
Vogler et al. (\cite{vogler}) 
for M~83 for the regions between spiral
arms.  For M~83, the continuum contributes no more than
$\approx$ 10\% of the total MIR 4 to 18 $\mu$m emission with
$\approx$ 90\% due to PAHs in these regions.  Although we do not have
CVF data for NGC~5907, all other indicators suggest that the situation
is likely comparable.  

It is quite clear that the PAH distribution follows that of the CO
(Fig.~\ref{major_axis}).  Both the $\lambda$850 $\mu$m and 
 $\lambda$1.2 mm cool dust also follow the CO distribution, and the
PAH and $\lambda$850 $\mu$m are spatially well correlated within
the disk of NGC~5907.
This suggests that both the PAHs and larger grains are associated with
the molecular gas component.  
PAHs are well known to trace 
classical photodissociation regions (PDRs) of molecular clouds which
are excited by far-UV photons from a hot stellar population in our
own Galaxy (see Sect.~\ref{spectrum}).
However
there is also evidence that the PAHs are extended radially
in comparison to CO both from our results as well as the cool 1.2 mm dust
distribution of Dumke et al. (\cite{dumke97}).
The cool dust and PAH emission both extend
to radii of $\sim$ 250$^{\prime\prime}$ in comparison to 
$\sim$ 175$^{\prime\prime}$ for CO. 
 While there may be some weak CO
present, the PAH and cool dust distributions show an excess with
respect to the CO at large radii in comparison to
smaller radii.   These
results suggest that, while PAHs are strongly associated with
the molecular gas, a more widespread distribution, likely associated
with cool HI, is also present
but at a much weaker level of emission. 

 Haas et al. (\cite{haas}) also find 
a correlation between 
PAHs and cold dust, the latter as measured by
the SCUBA $\lambda$850 $\mu$m continuum emission, in galaxies.  These authors 
suggest that PAH carriers are widespread and correlate with
large dust grains and neutral molecular clouds, but require a minimal
radiation field in order to be excited, thus explaining global
correlations between SFR and PAH emission as well as the fact that PAHs
are distributed more widely than the SF regions alone.  
This is also consistent with 
Peeters et al. (\cite{peeters04}) who suggest that PAHs may be better tracers
of B stars, which dominate the Galactic stellar
energy budget, rather than than as a tracer of massive star formation.
%Other independent evidence
%supports this view.  The [CII] 158 $\mu$m line is the major coolant for
%neutral HI (Hollenback \& Tielens 1999)
%and has been observed in an extended component associated with HI in NGC~6946
%(Madden et al. 1993).  At the same time, the
%[CII]/AFE ratio
%(the denominator meaning 'aromatic features seen in emission', corresponding
%to our designation of PAH features)
%has been found to remain 
%roughly constant over a wide range of SFR in many galaxies
%(Contursi et al. 2003, Helou et al. 2001), effectively linking [CII],
%PAHs and cool HI together.
 As indicated in Sect.~\ref{spectrum},
the PAHs in our Galaxy peak outside of HII regions but are also
profuse emitters in the diffuse ISM of galaxy disks, where the excitation 
may be from optical photons from an older population.
These studies therefore
support the association of PAHs with 
more extended neutral atomic gas.  
Tacconi-Garman et al. (\cite{tacconi-garman}) find 
that the {\it global} distribution of PAHs in NGC~253 and NGC~1808
follows the starburst activity, but on small scales, there is little
correspondence (correlation or anti-correlation) with star forming
regions.  It may well be that PAHs are widespread and excited by both
UV photons from a hot stellar population as well as optical photons
from the more widespread stellar population from which 
the interstellar radiation field (ISRF) originates.
The fact that they
correlate globally with the CO distribution would then be a statement
that there are more UV photons available at the appropriate wavelength
and also that there is a greater concentration of dust and PAHs in
regions of higher density molecular clouds.  In regions (or galaxies)
with very high SFRs
the PAHs will be destroyed in comparison to
the VSGs; this is supported by observations of M~83 
(Vogler et al. \cite{vogler})
and of starbursts and AGNS (e.g. Sturm et al. \cite{sturm00}).

An important result is the presence of PAH emission to very high latitude,
with scale heights of order 3.5 to 5 kpc and up to 6.5 kpc in
some locations.  By comparison, the
CO FWHM in the z direction is only $\sim$ 8$^{\prime\prime}$ = 426 pc
after deconvolution with the disk 
(Dumke et al. \cite{dumke97}) and the 
$\lambda$850 $\mu$m exponential scale height is only 110 pc to
the limits of those data.
An equally significant result is that such high latitude gas is seen
in a galaxy with a low SFR.   To our knowledge, 
this is the first evidence for high latitude
PAH emission in any external galaxy.  A  $\lambda$3.3 $\mu$m PAH spur of
$z$ extent $<$ 120 pc has now been
detected by 
Tacconi-Garman et al. (\cite{tacconi-garman}) in the outflowing wind of 
the starburst galaxy, NGC~253, lending support to the concept of 
'mass-loaded'
nuclear starburst winds and the possibility that some PAHs can 
survive such
blasts, the intense UV radiation and shocks that are expected to accompany
them. 

The situation is quite different in NGC~5907, however, which does not
show strong starbursting or outflowing winds.
 There is, however,
 a feature with anomalous velocities in 
high resolution  CO
data that cannot be explained by a bar or rotation model 
(Garcia-Burillo
et al. \cite{garcia-burillo97}, their Feature F).  
A CO loop-like feature 
of size $\sim$ 20$^{\prime\prime}$ associated with the anomalous
structure in the CO maps is suggestive
of a small-scale wind outflow. The radio point-source described in
Sect.~\ref{radio_correlation} (not at the same location as the CO feature)
indicates that some SF activity
is indeed occurring now in NGC~5907, although at a low level in
comparison to other SF galaxies.
The main independent evidence for high latitude emission 
in NGC~5907 is from the
radio continuum (Dumke et al. \cite{dumke00} and Fig.~\ref{lw6_radio}) 
which shows
a $\lambda$20 cm thick disk scale length of 1.5 kpc and a 
large scale field
of $\sim$ 1 $\mu$G.  This indicates that cosmic rays and magnetic fields
do exist in the halo, although a detailed search for a 
correlation has not 
yet been
possible (Sect.~\ref{radio_correlation}). 

A mechanism for depositing 
PAHs in the halo of a low SFR galaxy is not yet clear.  Since
SF and
magnetic fields are both clearly present in NGC~5907,
the PAHs may be vented into the halo initially from
 SNe and stellar winds from SF regions and
then assisted in reaching high latitudes
via coupling to the magnetic field.   Excitation could be
from UV photons that
leak from the disk into the halo.  An interesting possibility is that
the 3-D VSGs are themselves converted into planar 
 PAH macromolecules 
(Duley et al \cite{duley}).  This would occur in regions of shocks and high
UV excitation in the disk.  The smaller PAHs may then be more easily
'levitated' above the plane.  Jones et al. (\cite{jones}) have also noted that
PAHs and smaller particles can build up at the expense of larger ones
in shocks or winds and that wind velocities as low as tens of
km/s can modify the grain properties.  
The 
apparent dearth of $\lambda$850 $\mu$m emission in the halo
(Sect.~\ref{high_latitude_emission}) supports
this view. 
 It would be interesting to
obtain more sensitive $\lambda$850 $\mu$m and $\lambda$1.2 mm data 
in order to compare the results with starburst galaxies,
such as NGC~5775 in which 
$\lambda$850 $\mu$m emission has been detected in
the halo
 (Brar et al. \cite{brar03}).  Determining the admixture of different
dust components as a function of height would provide important clues
as to what mechanisms may be involved.
Conditions in galaxy halos
are quite different from those in the disk and may more closely
resemble those seen in elliptical galaxies.

Although there is a large optical ring around
NGC~5907, as described in Sect.~\ref{ngc5907}, there is no evidence 
that the high latitude
PAH emission is directly 
related to it.  The intersection of the optical ring
 with the disk of NGC~5907 occurs at two locations,
one of which (to the north) is not in the field of view of our maps 
and the other is near the southern tip of the major axis
(coordinates given in Fig.~\ref{lw10_dss}).
  We see
no evidence for enhanced emission or extensions at the latter location.
There does appear to be a disturbance in the H$\alpha$ map of Rand
et al. (\cite{rand}) approximately at this location, however.

%However, if we accept the model of Haas et al. (2002), 
%it is interesting to make some rather crude estimates.  They
%claim that the ratio F$_{7.7\,\mu m\,\,peak}$/F$_{850\,\mu m}\,\approx\,\,$2
%for galaxies.
%There is, as yet, no published 850 $\mu$m map of NGC~5907.   If the 
%Using the values for M~83 (Vogler et al. 2005), at the center of M~83,
%the peak 7.7 $\mu$m flux density is 1 Jy and the total for the LW6
%band (containing the 7.7 $\mu$m feature) is about 4 Jy.  Thus the ratio
%of peak to total in the band is $\sim$ 0.25.  Integrating over the observed
%high latitude emission on the east side only of the disk in our LW6 band
%(Fig.~\ref{lw6})
%we find a total flux density of 58 $\pm$ 8 Jy
%over a region 11300 square arcsec which suggests (using the same ratio
%as M~83)
%a peak 7.7 $\mu$m value of 15 Jy.  Using the Haas et al. correspondence to
%850 $\mu$m emission, this suggests that an 850 $\mu$m flux in the same
%region would be about 8 Jy. If we now use the conversion 
%to dust mass provide by ...
%\begin{equation}
%M_d\,=\,\frac{S_{850}\,D^2}
%{\kappa_d(\nu)\,B(\nu,\,T_d)}
%\end{equation}
%from Dunne et al. (2000), where
%$\kappa_d(\nu)\,=\,0.077$ m$^2$ kg$^{-1}$ is the dust mass opacity coefficient,
%$B(\nu,\,T_d)$ is the value of the Planck function at dust temperature, $T_d$,
%taken to be ... for NGC~5907
%I get 1.2 times 10^8 Msun which is too high.

%______________________________________________________________

\section{Conclusions}
\label{conclusions}

We have mapped the MIR emission of the low SFR galaxy, NGC~5907, in the
6.7W, 6.8N, 7.7N, 9.6N, 11.3N, and 12W  
(LW2, LW5, LW6, LW7, LW8, and LW10, respectively) ISO bands, 
at high sensitivity and high resolution. 
Although we do not have spectra
for NGC~5907, we have compared 
the line strengths and line ratios with those of other galaxies, 
especially
M~83 in quiescent regions between the spiral arms.  These 
comparisons indicate that the
excitation conditions are quite typical of regions and/or galaxies with
a low SFR.  The contribution from hot VSGs is likely no more than $\sim$
10\% across these bands and is likely negligible in all bands except
the broad, IRAS-equivalent 12W band.  The MIR emission is dominated
by PAH emission.

In the disk, the high spatial resolution (1$^{\prime\prime}$)
broad band 12W emission is peaked towards the optical
dust lane.  The bands dominated by PAH emission (at 6$^{\prime\prime}$
resolution) are observed all along the
major axis of the galaxy and extend
almost to the end of the optical disk (cf $\sim$ 35 kpc in comparison
to 41 kpc) and possibly farther.  The distribution of PAHs in the disk
roughly follows that of the CO and not the broad HI distribution.
The PAH distribution shows an apparent correlation with the
SCUBA $\lambda$850 $\mu$m distribution to within the positional
errors of the data.   The $\lambda$1.2 mm emission also follows
that of the molecular gas.  At the same time, there is some PAH emission
(and also $\lambda$1.2 mm emission)
at larger galactocentric radii than the CO, suggesting that PAH
emission is more widespread than the molecular gas alone.  It is
likely
that greater PAH emission is seen where there is more
molecular gas because there are more UV photons available from SF in
these regions and because of the greater concentration of grains and
PAHs in higher density gas.  A smaller fraction of the PAH emission
will be coming from the more extended HI, excited by the general
diffuse ISRF.

An important finding of this work is the detection of PAH emission
at high latitudes.  The emission features show structure
in the narrow bands that isolate single PAH features, appearing similar
to features seen in H$\alpha$, radio continuum, and other bands in
other edge-on galaxies (see, e.g. Lee et al. \cite{lee02}).  Individual features
extend to latitudes of up to 6.5 kpc but a more typical scale height is
3.5 kpc.  Previous high latitude emission on kpc scales has been 
seen only in
the radio continuum component in this galaxy.  Thus, 
cosmic rays and magnetic
fields are known to exist in the halo, but other components
have not (yet) been observed.  Indeed, the PAHs appear to be
selectively represented in the halo in comparison to large
grains.  We note also an
increase in the $\lambda$11.3/$\lambda$7.7 PAH ratio with distance 
from the major axis.
Although NGC~5907 is a low SFR galaxy, there is some SF 
activity occurring,
such as has been seen via some anomalous CO emission as well as a 
previously unreported off-nucleus radio continuum point source which 
is likely a supernova remnant or remnants.  Thus, the halo PAHs may 
be transported to high latitudes 
with the help of coupling to magnetic fields 
(likely with gas as well), from regions of SF.

Although NGC~5907 is surrounded by a faint stellar ring, there is no
indication in these data that interaction with the ring plays a 
direct role
in the presence of the high latitude emission.

%______________________________________________________________

\begin{acknowledgements}
     We are very grateful to Pierre Chanial for developing, at short notice,
a seamless data reduction package for beam-switched observations.  Many
thanks also to Fr{\'e}d{\'e}ric Galliano for his assistance with the process of ISO
data reduction.  Thanks to Elias Brinks for making the HI cube
of this galaxy available for our use and to Manoulis Xilouris for
providing the SCUBA 850 $\mu$m data.  We wish to thank Herv{\'e}
Aussel also for providing software and assisting with checking
the quality of the data and to Matthew Ashby for helpful suggestions.
JAI wishes to thank the director, administration and staff of
CEA/Saclay for hospitality during a sabbatical leave.  This work has
been supported by the Natural Sciences and Engineering Research
Council of Canada.
\end{acknowledgements}

\end{document}